# Search for binary companions around millisecond pulsars

Arpita Choudhary
*Madras School of Economics, Chennai, India*

24 July 2025

## 1 INTRODUCTION

It is believed that the radio pulsars rotating at spin periods of about 30 millisecond or even lower got such high spins through the transfer of angular momentum by accreting matter from their binary companions (or past companions) in the past. That is why they are often called as 'recycled pulsars' and most of the time we refer to them as millisecond pulsars (MSPs). For a long time, only support for this recycling scenario (Alpar et al. 1982; Radhakrishnan & Srinivasan 1982) was the fact that most of the MSPs are in binary systems. The recent discovery of millisecond pulsars that swing between an accretion powered (X-ray) and a rotation powered (radio) state (Archibald et al. 2009; Papitto et al. 2013; Roy et al. 2015) established this theory beyond doubt.

However, around 35% of MSPs are isolated, and 50% of those are in globular clusters where dynamical interactions can be responsible for snapping the binaries after the recycling phase. For the rest 50% of isolated MSPs, located either in the Galactic disk or in the halo, the most likely scenario is the disruption of binaries during the second supernova in systems where even the secondary stars were massive enough.

Some of the isolated pulsars might actually be in binaries - either not timed very well to obtain the signature of the orbit or in extremely wide binaries that can be revealed only through very accurate timing, as in the case of PSR J1024−0719 (Bassa et al. 2016; Kaplan et al. 2016).

Interestingly, more than a decade ago, Sutaria et al. (2003) found the presence of two stars in optical bands near the position of PSR J1024−0719. But due to astrometric uncertainties, they could not draw any firm conclusions. Timing analysis of PSR J1024−0719 for long enough time base-line recently revealed that the brighter one among the two seen by Sutaria et al. (2003) is actually a gravitationaly bound companion of PSR J1024−0719 Bassa et al. (2016).

This has motivated us to pursue an exercise to identify probable candidates of companions of MSPs in the Galactic field. This exercise has even the potential of discovering optical emission from MSPs themselves, which was the goal of Sutaria et al. (2003).

Discovery of any possible binary companion will be followed by motivated pulsar timing campaigns to confirm the binary nature and to constrain orbital parameters.

## 2 SELECTION CRITERIA

As we have already mentioned, dynamical interactions can lead to the formation of isolated MSPs in globular clusters, especially if the initial binary was wide enough (Bagchi & Ray 2009). So it is unlikely that the isolated MSPs in the globular clusters would be part of very wide binaries. Moreover, the crowded fields of clusters would make the identification of probable companion difficult. Thus we exclude isolated MSPs in clusters from our study.

We also use the loose definition of MSP, i.e. spin period being less than 30 ms. There are 60 such pulsars listed in the ATNF catalogue[1] (Manchester et al. 2005).

## 3 SEARCHING FOR COMPANIONS IN DIFFERENT IMAGING ARCHIVAL DATA

Sutaria et al. (2003) performed photometry and astrometry on pulsars PSR J1024−0719 and J1744−1134. Comparing the radio timing positions and VLA positions, Sutaria et al. (2003) found a mean scatter of about 0.8″ in each coordinate (RA and DEC). After adding this to the astrometic errors, they found the total astrometric error in both RA and DEC would be ≈ 0.9″ giving an error circle of radius 1.3″ for both the pulsars. So, they considered any optical star-like emission within 2″ of the pulsar radio position as that of interest. This transfer to the projected separation $r = d\theta$ = 0.011829 pc when we use $d$ = 1.22 kpc, measured in recent timing analysis of the pulsar J1024−0719 (Bassa et al. 2016). We have also decided to set the limit of search as $r_{max}$ = 0.0118 pc. We calculate the search limit for each pulsar as $\theta_{max} = r_{max}/d$ based on the known distances of the pulsars.

We first downloaded the DSS images of size 5 ′ centered at the positions of those pulsars. Afterwards, we manually checked all these DSS images using ds9 package. Our companion search circle limit was not rigid, we keep bright candidates even if they are few arcseconds farther than $\theta_{max}$ of that particular pulsar. We find 15 best candidates, for which details are given in Table 1. Archival images for these sources can be found in Appendix-A.

---

[1] http://www.atnf.csiro.au/research/pulsar/psrcat/expert.html

Table 1. Isolated MSPs in the Galactic field with probable companions. The parameters in the table from left to right are serial no, Jname of the pulsar, the right ascension (RA), the declination (DEC), the distance of the pulsar from the earth ($d$), the angular radius of the search circle ($\theta_{max}$), the angular distance of the probable companion from the pulsar ($\theta_{cand}$), the separation between the pulsar and the companion if the optical star is the companion of the pulsar ($r_{cand} = d\theta_{cand}$).

| S.No. | Name | RA (hh:mm:ss) | DEC (dd:mm:ss) | References Coordinates | $d$ (kpc) | $\theta_{max}$ | $\theta_{cand}$ | $r_{cand}$ (pc) |
|---|---|---|---|---|---|---|---|---|
| 1 | J0030+0451 | 00:30:27.42826(5) | +04:51:39.711(2) | Matthews et al. (2016) | 0.34 | 7.28″ | 3.2″ & 7.19″ | 0.005 |
| 2 | J0711−6830 | 07:11:54.189114(13) | -68:30:47.41446(8) | Reardon et al. (2016) | 0.11 | 22.5″ | 7.98″,16.39″... | 0.004,0.008... |
| 3 | J0740+6620 | 07:40:45.798(5) | +66:20:33.65(2) | Guillemot et al. (2016) | 0.93 | 2.66″ | 4.86″ | 0.02 |
| 4 | J0922−52 | 09:22:00 | -52:00:00 | Mickaliger et al. (2012) | 0.35 | 7.07″ | 6.3″ & 8.7″ | 0.010,0.014 |
| 5 | J1536−4948 | 15:36:24.016 | -49:48:45.39 | Roy et al. (2012) | 0.98 | 2.52″ | 3.6″ | 0.017 |
| 6 | J1546−59 | 15:46:00 | -59:00:00 | Mickaliger et al. (2012) | 3.89 | 0.63″ | 1.1″ | 0.020 |
| 7 | J1652−48 | 16:52:9(5) | -48:45(7) | Knispel et al. (2013) | 4.39 | 0.56″ | 1.1″ | 0.023 |
| 8 | J1658−5324 | 16:58:39.34359(9) | -53:24:07.003(1) | Camilo et al. (2015) | 0.88 | 2.81″ | 2.3″ | 0.009 |
| 9 | J1730−2304 | 17:30:21.66624(8) | -23:04:31.19(2) | Reardon et al. (2016) | 0.62 | 3.99″ | 2.1″ & 3.6″ | 0.006,0.01 |
| 10 | J1744−1134 | 17:44:29.407190(3) | -11:34:54.6925(2) | Matthews et al. (2016) | 0.39 | 6.26″ | 2.7″ | 0.005 |
| 11 | J1801−1417 | 18:01:51.073331(19) | -14:17:34.526(2) | Desvignes et al. (2016) | 1.1 | 2.25″ | 4.7″ & 5.9″ | 0.025,0.031 |
| 12 | J1843−1448 | 18:43:01.375(3) | -14:48:12.61(3) | Lorimer et al. (2015) | 3.47 | 0.71″ | 1.06″ | 0.017 |
| 13 | J1902−70 | 19:02:00 | -70:00:00 | Ray et al. (2012) | 0.92 | 2.69″ | 4.3″ | 0.019 |
| 14 | J1905+0400 | 19:05:28.273436(16) | +04:00:10.8830(6) | Gonzalez et al. (2011) | 1.7 | 1.45″ | 2.4″ & 3.1″ | 0.019,0.025 |
| 15 | B1937+21 | 19:39:38.561227(2) | +21:34:59.12567(5) | Matthews et al. (2016) | 3.5 | 0.70″ | 2.2″ | 0.037 |

## 4 RESULTS OF SHORTLISTED CANDIDATES

For each of the fifteen shortlisted pulsars, we explored various telescope data archives at different wavelengths e.g. HST, UKIDSS, Spitzer (IRAC-MIPS), VLT, Herschel, etc. and found the imaging data listed in Table 2. The Galex, Swift and PS1 fields are not deep enough to resolve the candidates in all cases.

### 4.1 PSR J0030+0451

Becker et al. (2000) studied the pulsar in X-ray regime. Optical observations with VLT were done by Koptsevich et al. (2003). They could not detect any optical counterpart to the pulsar down to B>27.3, V>27.0 and R>27.0 in the immediate neighborhood. The nearest detected source was 3″ away, but was rejected on astrometric grounds. Lommen et al. (2005) did the parallax and proper motion measurement of the pulsar using data from Arecibo telescope. Apart from the source at 3″ distance from the pulsar, we detected another bright 2MASS source in the VLT image (Fig A1), 7.19″ away from the pulsar position having B=18.2. This object has not been investigated yet.

### 4.2 PSR J0711-6830

This pulsar is located at a distance of 0.11 kpc from us, and is the nearest candidate in our list. There is no publicly available archival imaging data for this pulsar. Searching NED catalogue for pulsar neighborhood, we see few stars in the DSS image (See Fig A2). The details of the sources are listed in table 3. Out of the mentioned five, source number 2 shows a proper motion of 14 ± 1 in R.A. and −38 ± 32 in declination.

We have checked NASA-ADS for papers on this pulsar. We found a couple of papers, e.g., Bailes et al. (1997), and Hotan et al. (2006). But none of these two searched in the imaging data for companions.

### 4.3 PSR J0740+6620

This pulsar at a distance of 0.93 Kpc from us was studied by Stovall et al. (2014) using the Green Bank telescope data. They gave initial discovery parameters of the pulsar. Two years later, Guillemot et al. (2016) did the proper motion and distance measurement studies of various pulsars and for J0740+6620, they discovered significant GeV $\gamma$-ray signals. No archival optical/IR data is available for this pulsar so far. In the DSS cut-out image (Fig A3), we found an object located 4.86″ away from the pulsar position, having B=21.47 and R=16.67.

### 4.4 PSR J0922-52

Using the archival data from the Parkes Multibeam Pulsar Survey, Mickaliger et al. (2012), made discovery of this MSP. No research has been done on finding its optical counterparts/companions. We found archival Spitzer-IRAC data for this MSP, but the observations made were not deep enough to resolve the sources around pulsar position. In the DSS cut-out image (See Fig A4), we found two sources at distances 6.3 and 8.7″ away from the pulsar. The one closer to the pulsar position has J=16.08,H=15.23 and K=14.93. The other source has B=19.66 and R=18.48.

### 4.5 PSR J1536-4948

This pulsar was discovered by Roy et al. (2012) and later García et al. (2015) studied the X-ray properties of the pulsar using XMM-Newton telescope. There is no archival optical/IR imaging data available for this pulsar. In the DSS image (Fig A5), the optical NOMAD catalog suggests the presence of another object at a distance of 3.6″ from pulsar position. This object has B=19.63 and R=17.76. The object also has a proper motion in both R.A and Dec. as -30 and -16 respectively. No companion search has been done for this pulsar so far.

**Table 2.** Available archival imaging data

| S.No. | Pulsar | VLT | UKIDSS | Spitzer | Galex | Swift | PAN-STARRS(PS1) |
|---|---|---|---|---|---|---|---|
| 1 | J0030+0451 | yes | yes | yes | yes | - | - |
| 2 | J0711-6830 | - | - | - | - | yes | - |
| 3 | J0740+6620 | - | - | - | yes | - | - |
| 4 | J0922-52 | - | - | yes | - | - | - |
| 5 | J1536-4948 | - | - | - | - | yes | - |
| 6 | J1546-59 | - | - | - | - | - | - |
| 7 | J1652-48 | - | - | - | yes | - | - |
| 8 | J1658-5324 | - | - | - | - | yes | - |
| 9 | J1730-2304 | - | - | yes | yes | - | - |
| 10 | J1744-1134 | yes | - | yes | yes | - | - |
| 11 | J1801-1417 | - | - | - | - | - | - |
| 12 | J1843-1448 | - | yes | - | yes | - | - |
| 13 | J1902-70 | - | - | - | yes | - | - |
| 14 | J1905+0400 | - | yes | - | - | yes | - |
| 15 | B1937+21 | - | yes | yes | - | yes | - |

**Table 3.** List of probable companions around PSR J0711-6830. The serial numbers indicate the objects as shown in Fig A2.

| S.No. | Distance from pulsar (seconds of arc) | Magnitudes |
|---|---|---|
| 1 | 7.98 | B=20.36, R=19.95 |
| 2 | 16.39 | B=21.06, R=19.89 |
| 3 | 17.7 | - |
| 4 | 20.4 | R=20.19 |
| 5 | 22.9 | B=18.38, V=17.74, R=17.23, J=16.27, H=15.55, K=15.42 |

### 4.6 PSR J1546-59

Mickaliger et al. (2012) discovered this pulsar as well, using data from Parkes Multibeam Pulsar Survey. In absence of any archival optical/IR imaging data, we used DSS cut-out image (Fig A6) to find a source at distance 1.1″ from the pulsar having B=18.38, R=16.41, J=15.86, H=15.17 and K=15.16. The object also has a proper motion of 28 ± 51 in R.A. and −168 ± 42 in declination.

### 4.7 PSR J1652-48

In search for radio pulsars in compact binary systems in Parkes Multibeam Pulsar Survey, Knispel et al. (2013) discovered this pulsar to be part of a binary system, but no further investigation has been done on the nature of the binary. In the DSS image (See Fig A7), we mark the companion source shown in the NOMAD catalogue at a distance of 1.1″ from the pulsar with R=18.81.

### 4.8 PSR J1658-5324

In a survey of 14 unidentified Fermi Large Area Telescope sources in the southern sky, Kerr et al. (2012) discovered the MSP. In search of unidentified Fermi Gamma ray sources, Camilo et al. (2015) detected J1658−5324 as one. There is no publicly available imaging archival data for this pulsar. We found a source very close to the pulsar position in the DSS image (Fig A8) with B=18.7 and R=18.2. No companion search around this pulsar has been conducted so far.

### 4.9 PSR J1730-2304

Discovery of this southern hemisphere pulsar was made by Lorimer et al. (1995), who also categorized it as solitary. After that, no companion search around this pulsar was done. The first significant parallax measurements for the MSP was done by Reardon et al. (2016).

We found Spitzer-IRAC archival imaging data for this pulsar, but it is not deep enough to clearly resolve the nearest sources. In the DSS image (Fig A9), we mark the two nearest sources as 1 and 2, listed in the NOMAD catalogue. Source 1 at a distance of 3.6″ from the pulsar, has B=18.87, V=17.15, R=16.93, J=12.99, H=11.909 and K=11.59. The other source numbered 2 is much closer in position to the pulsar at 2.1″ and has R=18.88. This source also has proper motion of −16 ± 17 in R.A. and −40 ± 14 in declination.

### 4.10 PSR J1744-1134

Sutaria et al. (2003) gave an upper limit to any optical companion visibility near this pulsar. According to them, the faintest object that could be detected would have magnitudes: B=28.0, V=27.25 and R=26.8 for a $4\sigma$ detection over entire point spread function. In the archival VLT images, we see an object close to the star (Fig A10), but nothing could be detected in archival Spitzer-IRAC data. The object is also not listed is any catalogue so far.

We have checked NASA-ADS for papers on this pulsar. We found the pulsar discovery paper (Bailes et al. 1997). Sutaria et al. (2003) worked with VLT imaging data but could not reveal any companion to the pulsar.

### 4.11 PSR J1801-1417

Faulkner et al. (2004) again using the Parkes Multibeam Pulsar Survey made discovery of this millisecond pulsar and categorized it as a solitary pulsar and no further companion search was conducted on this object.

There is no archival imaging data available for this pulsar. The two nearest sources to the MSP as shown by the NOMAD catalogue as marked as 1 and 2 in the DSS cut-out image (Fig A11). Source 1 at 5.9″ from the MSP

has B=19.67, V=17.93, R=17.63, J=15.23, H=14.47 and K=14.26. The other source 2 at 4.7″ has J=15.99, H=14.98 and K=14.79.

### 4.12 PSR J1843-1448

This pulsar was discovered in the Parkes 20-cm multibeam pulsar survey of the Galactic plane and Lorimer et al. (2015) presented the timing observations of the MSP, designated as isolated. No further search for companion has been done on the same.

We found archival UKIDSS observations for this pulsar. The color composite image (Fig A12) shows an object very close to the pulsar at 1.06″, having B=19.3 and R=17.68. This object also has a proper motion of 20 ± 10 in R.A. and −8 ± 2 in declination.

### 4.13 PSR J1902-70

Ray et al. (2012) discovered this MSP while searching for over 300 LAT gamma-ray sources that do not have strong associations with known gamma-ray emitting source classes and have pulsar-like spectra and variability characteristics.

There is no available archival imaging data for this pulsar. So, we used DSS cut-out image to point out the nearest source to the MSP (Fig A13) at a distance of 4.3″ from the pulsar position with B=20.79 and R=19.66.

### 4.14 PSR J1905+0400

Discovery of this MSP was made by Hobbs et al. (2004) using data from the Parkes multibeam pulsar survey. A few years later, the first high precision timing measurements of the pulsar were conducted by Gonzalez et al. (2011). No companion search has been done for this pulsar.

We downloaded the UKIDSS archival imaging data available in JHK bands. The color image (Fig A14) shows the two nearest objects to the pulsar at 2.4 and 3.1″ distances, having R=17.05 and R=16.76 respectively.

### 4.15 PSR B1937+21

Infrared observational data is available for the first ever discovered pulsar, which is still considered isolated. In the archival Spitzer-IRAC/MIPS and UKIDSS data, we identify a star at the pulsar position (Fig A15), which is NSV 24840, a variable star. No optical observations have been recorded for it so far.

We have checked NASA-ADS for papers on this pulsar. We found several papers, e.g., Dolch et al. (2017), Ng et al. (2014), and Walker et al. (2013). But none of these searched in the imaging data for companions or counterparts .

## 5 FURTHER OBSERVATIONS

In our search in different archival imaging data (optical/Infrared/UV), we find bright sources near the radio positions of 16 pulsars listed in Table 1. But we can not confirm whether these are binary companions or just background/foreground stars. Better quality data will lead us to better understanding of the surroundings of these pulsars.

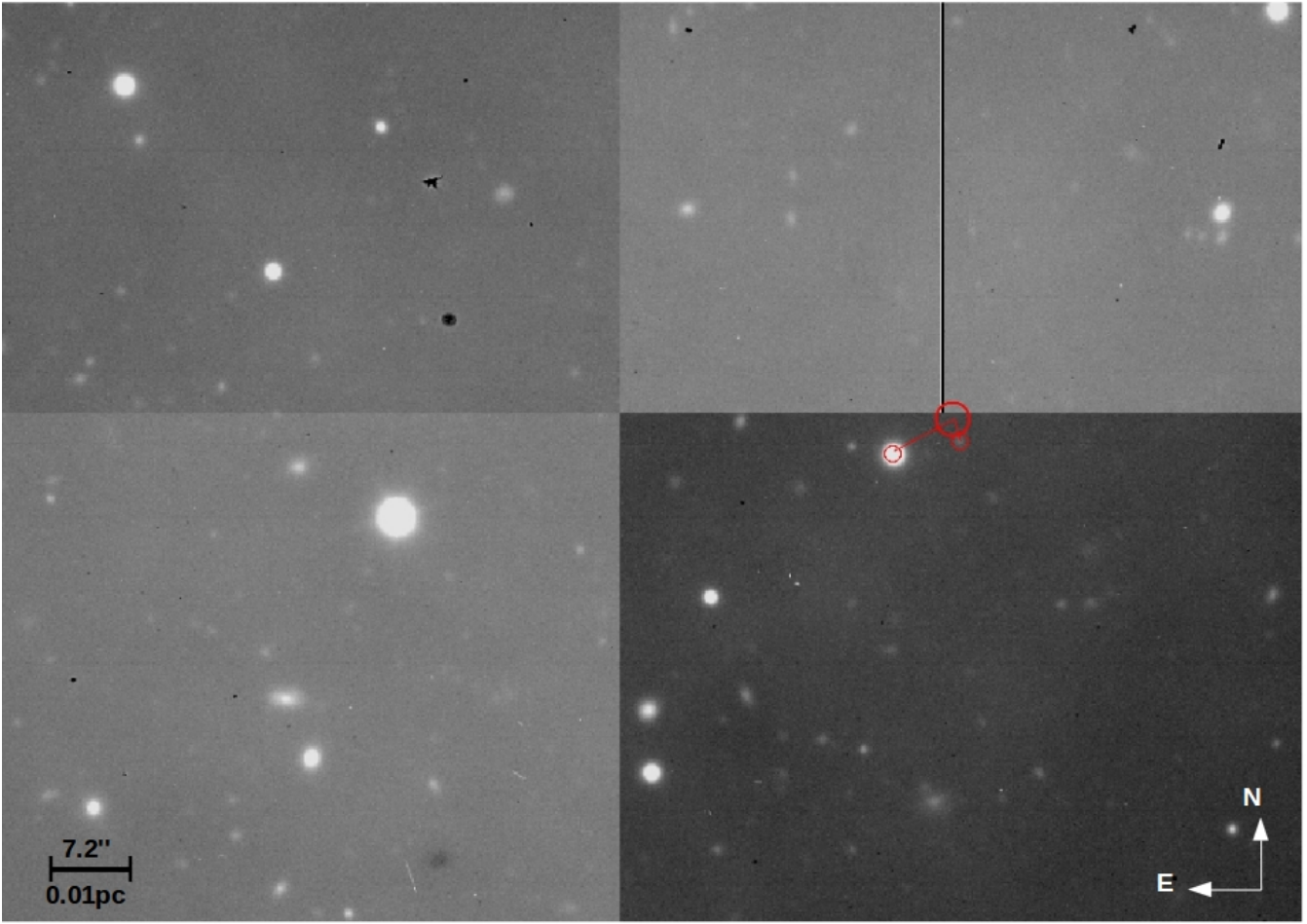

**Figure A1.** VLT R band image of PSR J0030+0451 showing pulsar position marked in big red circle and the stars marked in small red circles.

## APPENDIX A: ARCHIVAL IMAGES

This paper has been typeset from a T<sub>E</sub>X/LAT<sub>E</sub>X file prepared by the author.

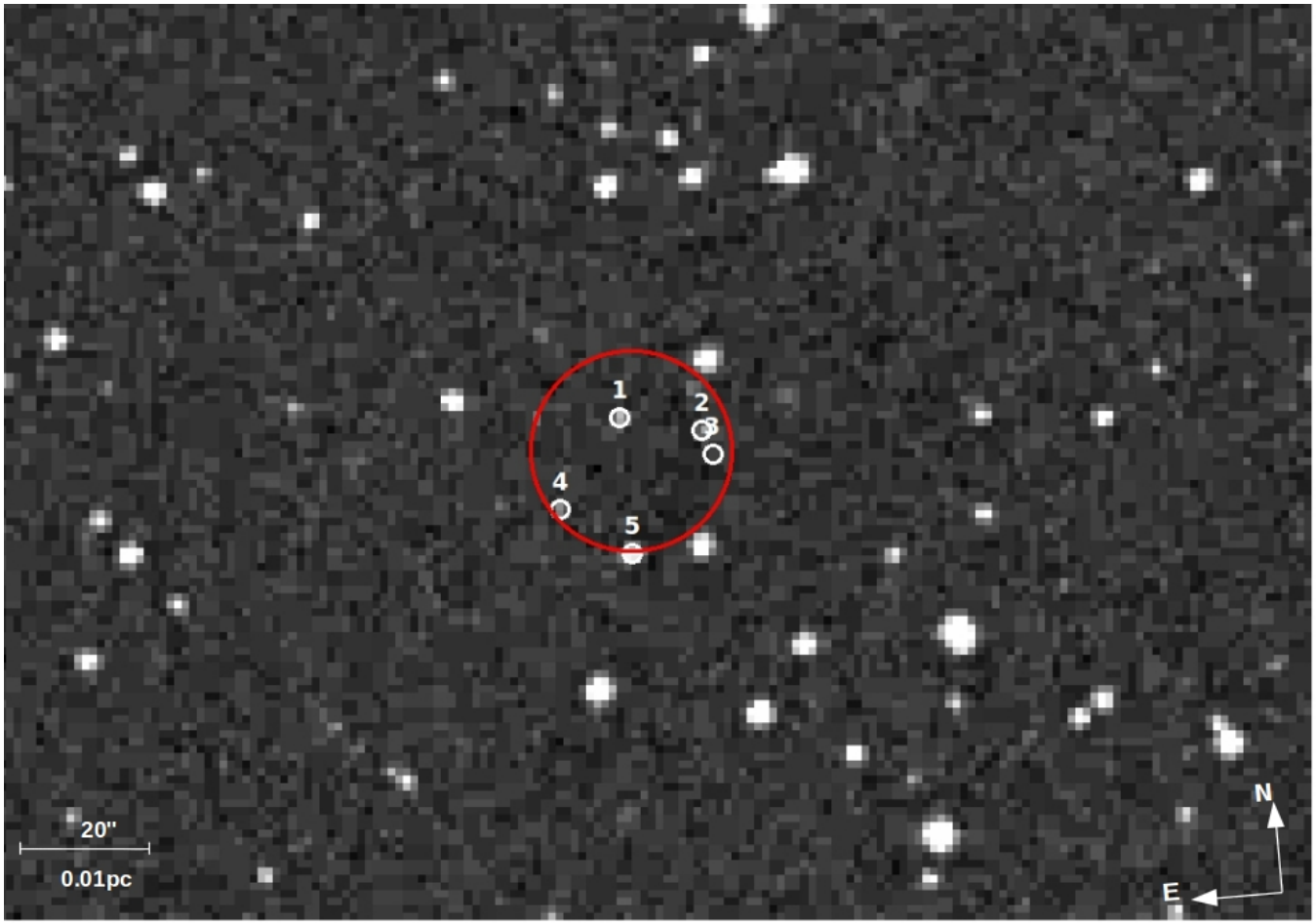

**Figure A2.** DSS image of PSR J0711-6830 showing pulsar position marked in red circle and the stars marked in white circles.

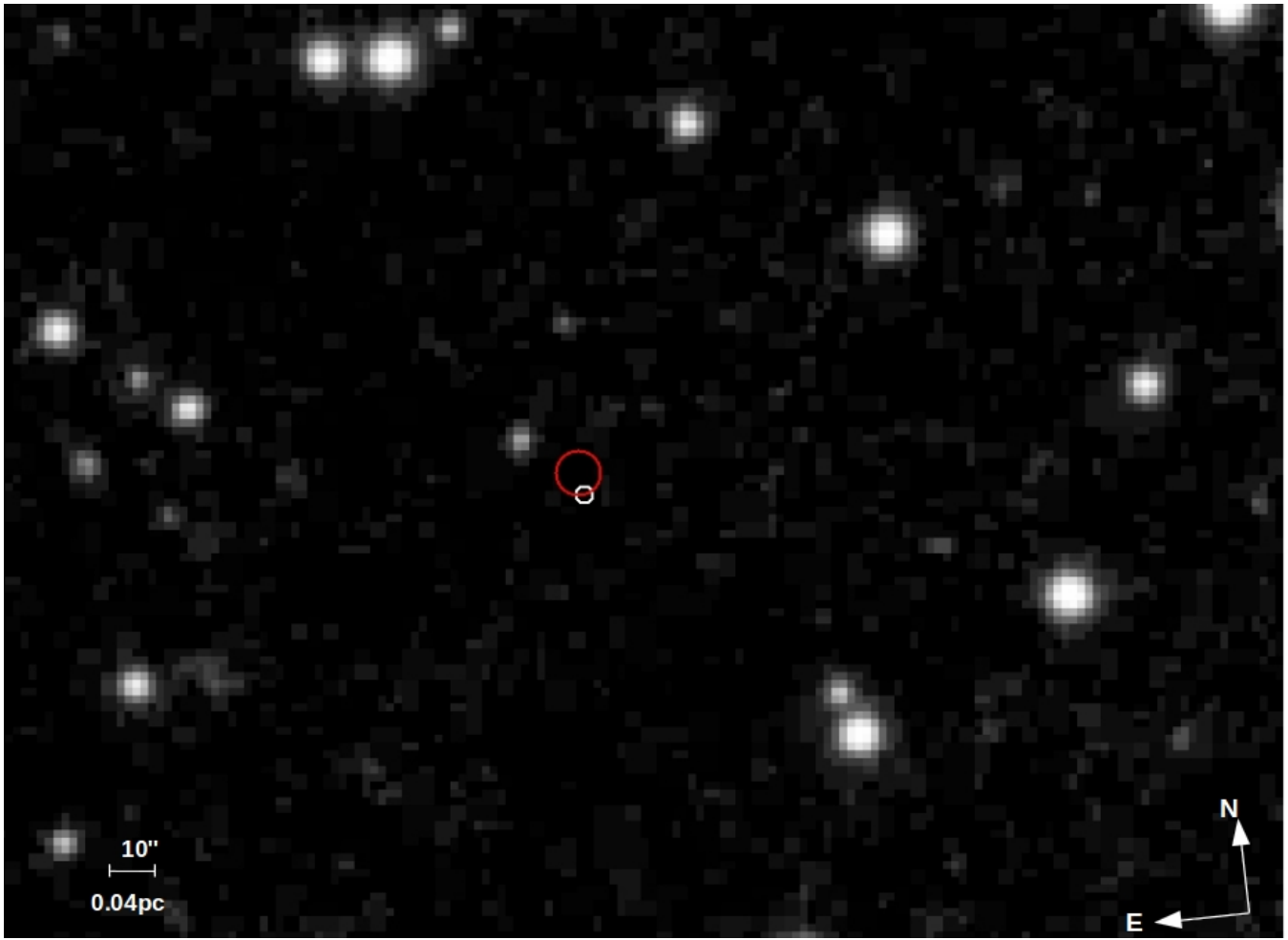

**Figure A3.** DSS image of PSR J0740+6620 showing pulsar position marked in red circle and the star marked in white circle.

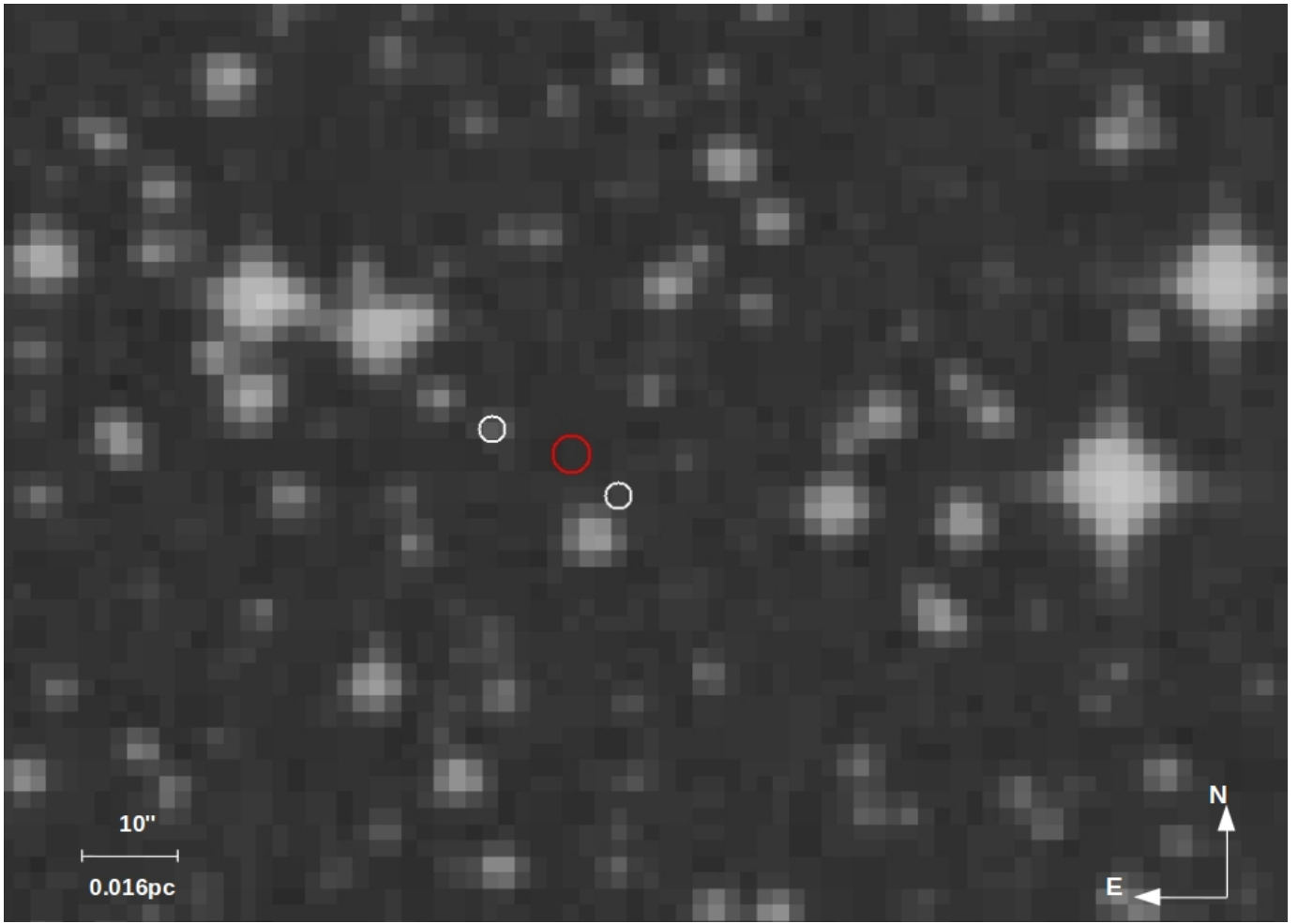

**Figure A4.** DSS image of PSR J0922-52 showing pulsar position marked in red circle and the stars marked in white circles.

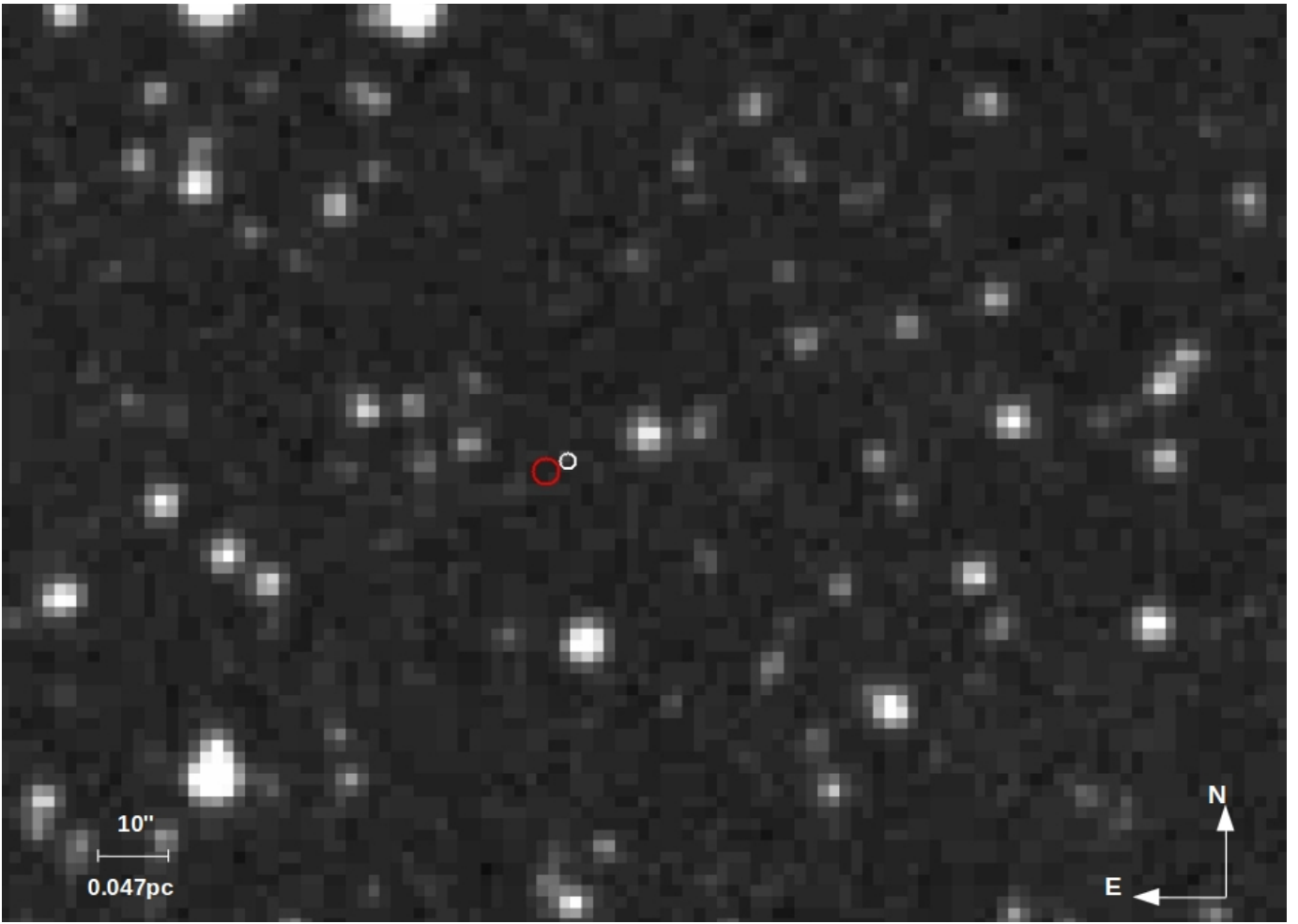

**Figure A5.** DSS image of PSR J1536-4948 showing pulsar position marked in red circle and the star marked in white circle.

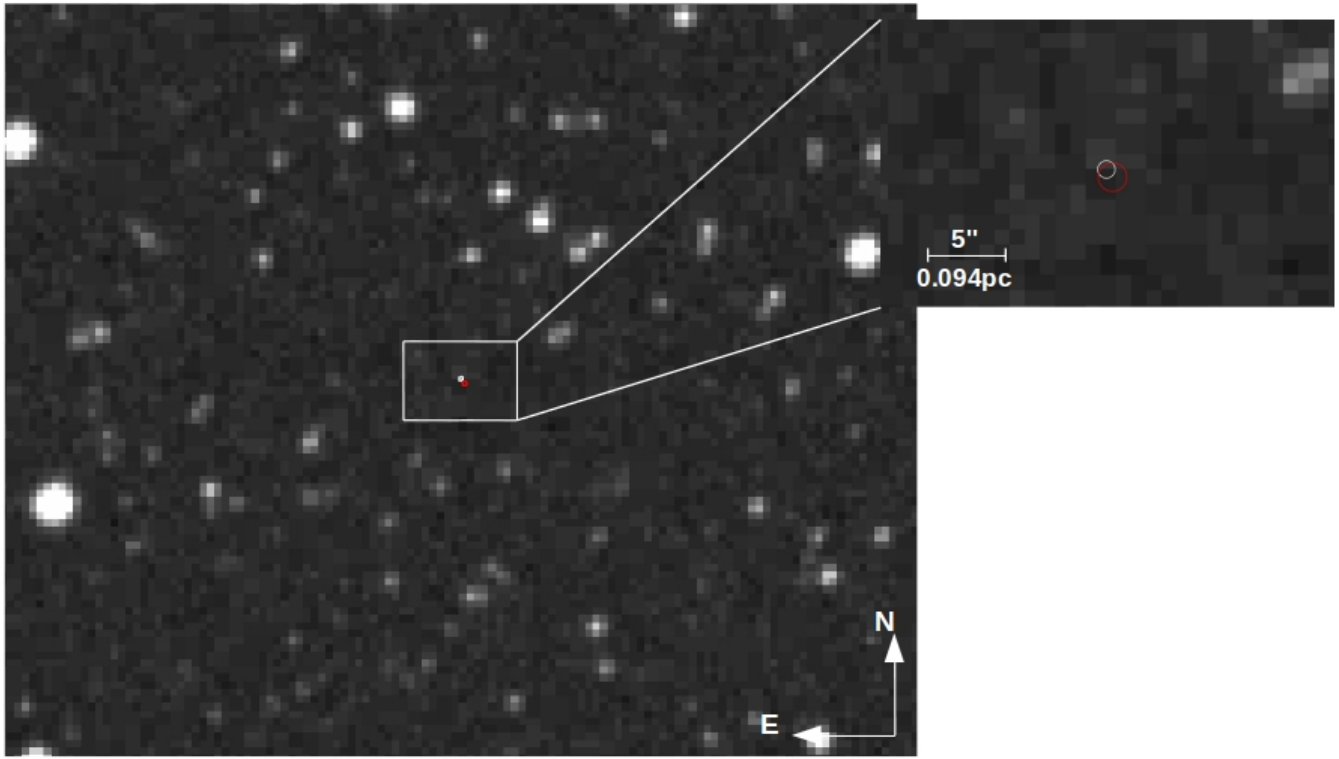

**Figure A6.** DSS image of PSR J1546-59 showing pulsar position marked in red circle and the star marked in white circle.

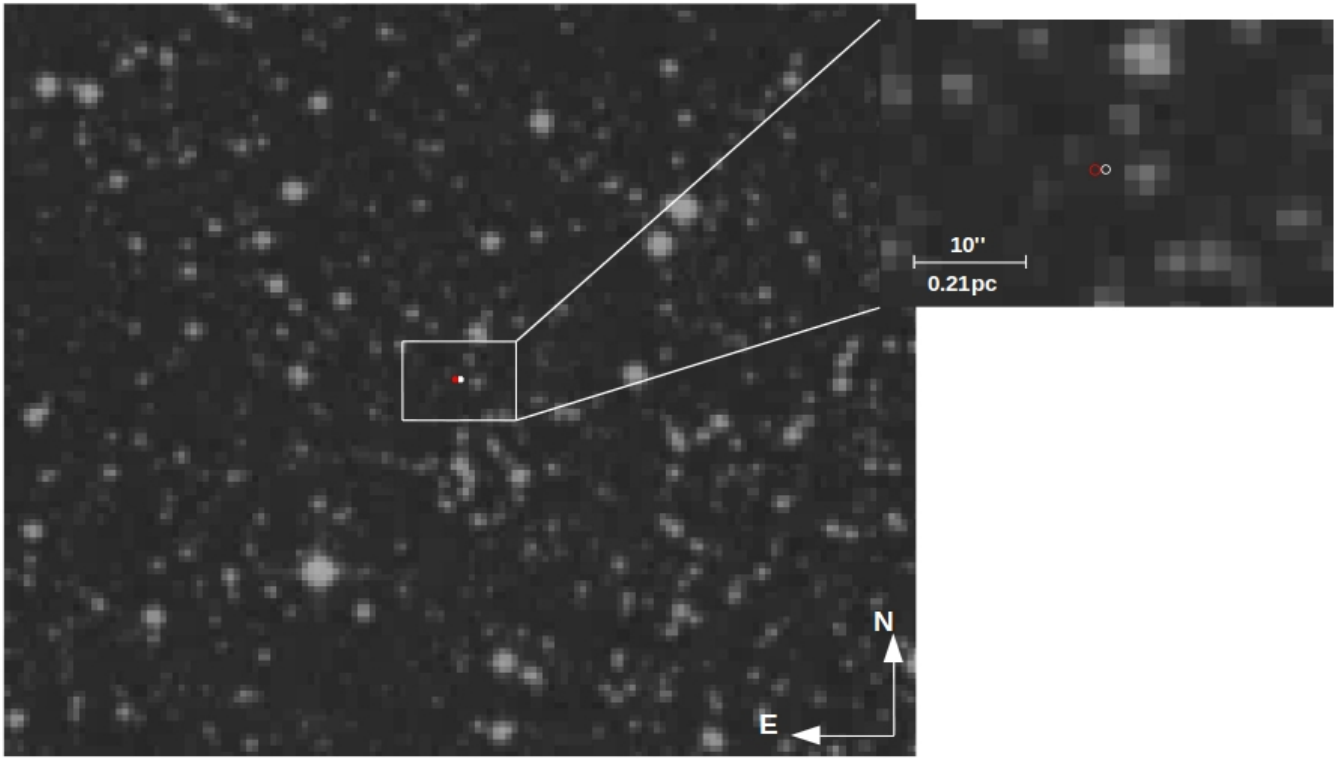

**Figure A7.** DSS image of PSR J1652-48 showing pulsar position marked in red circle and the star marked in white circle.

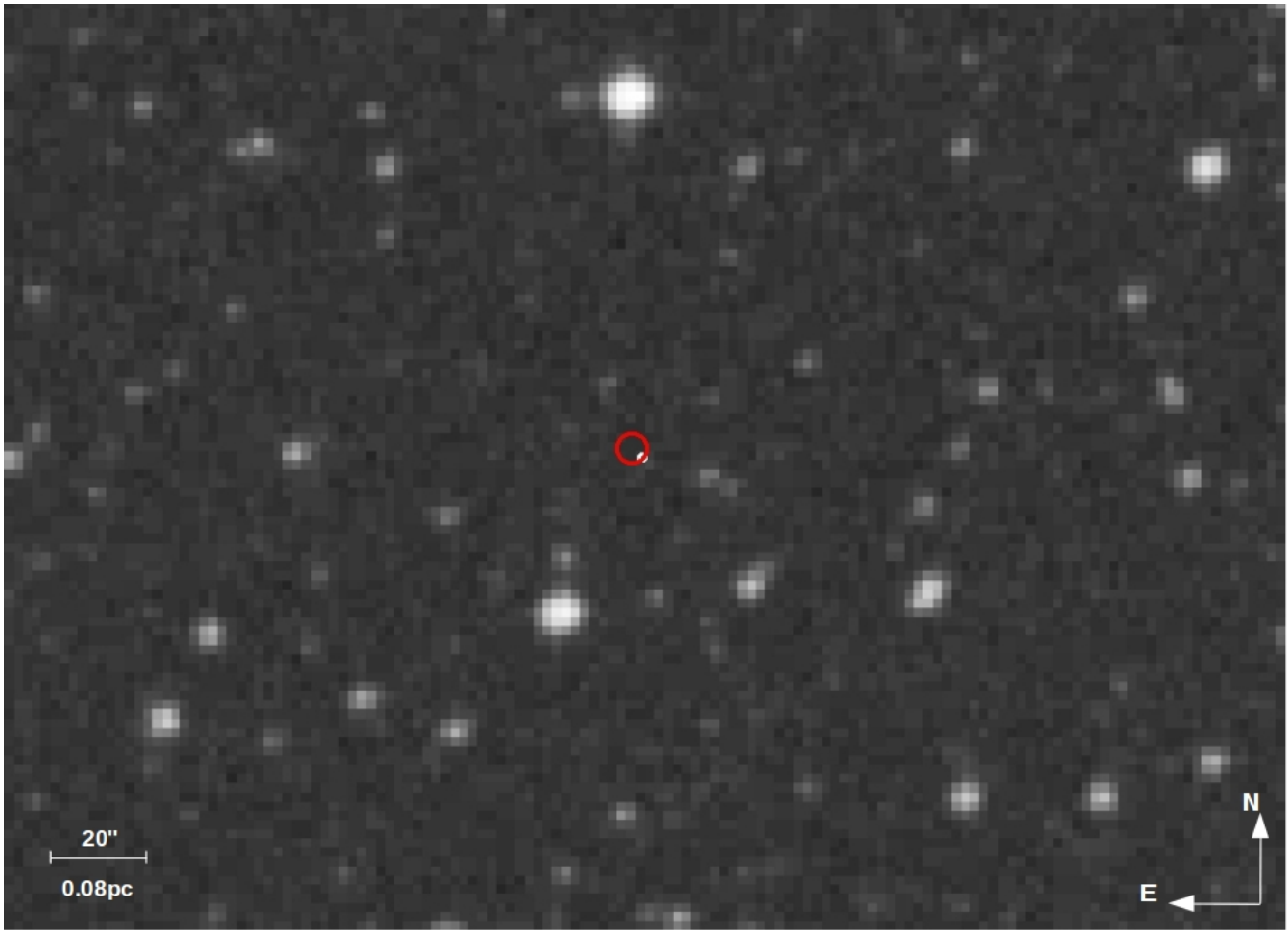

**Figure A8.** DSS image of PSR J1658-5324 showing pulsar position marked in red circle and the star marked in white circle.

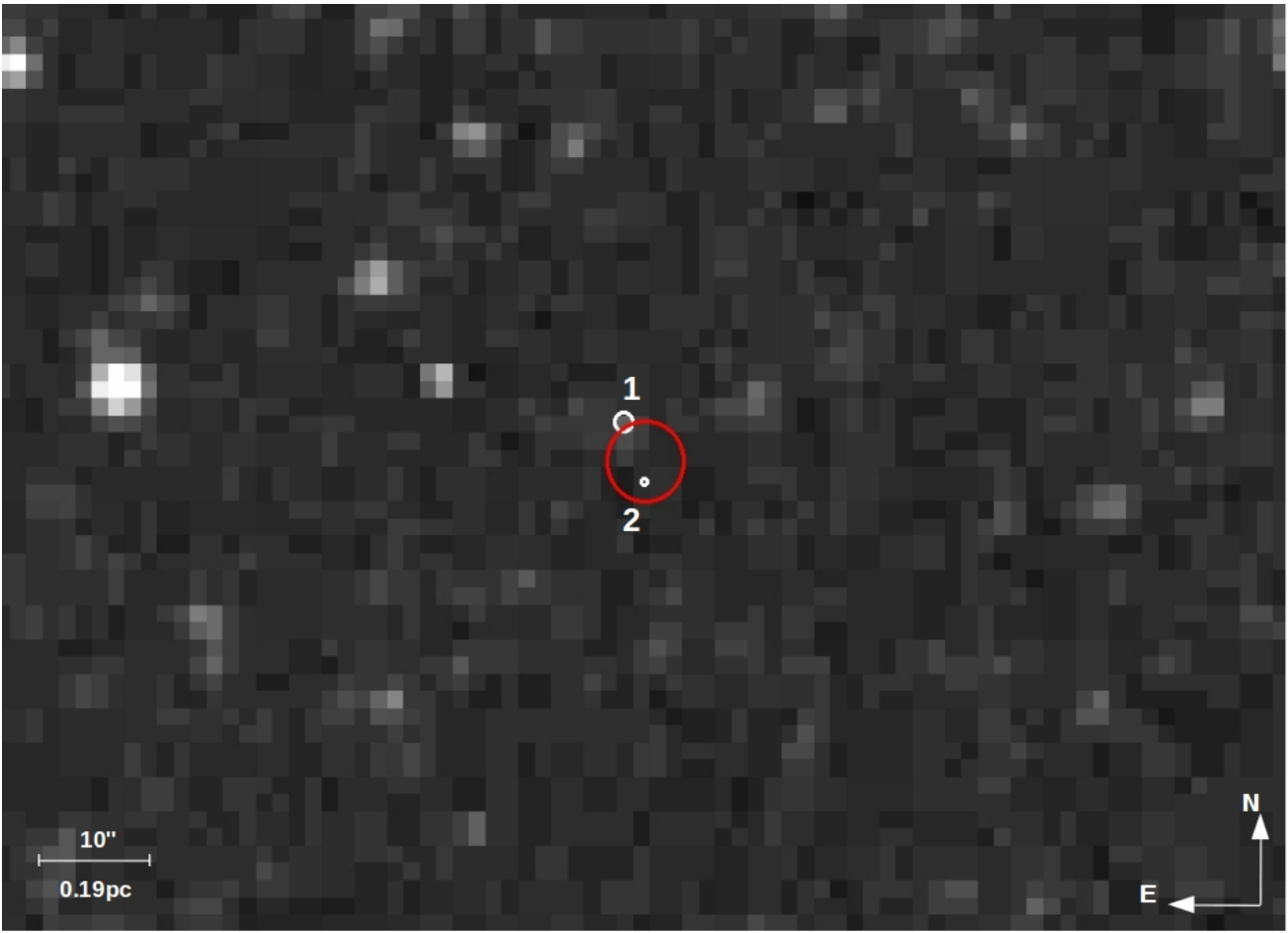

**Figure A9.** DSS image of PSR J1730-2304 showing pulsar position marked in red circle and the stars marked in white circles, numbered 1 and 2.

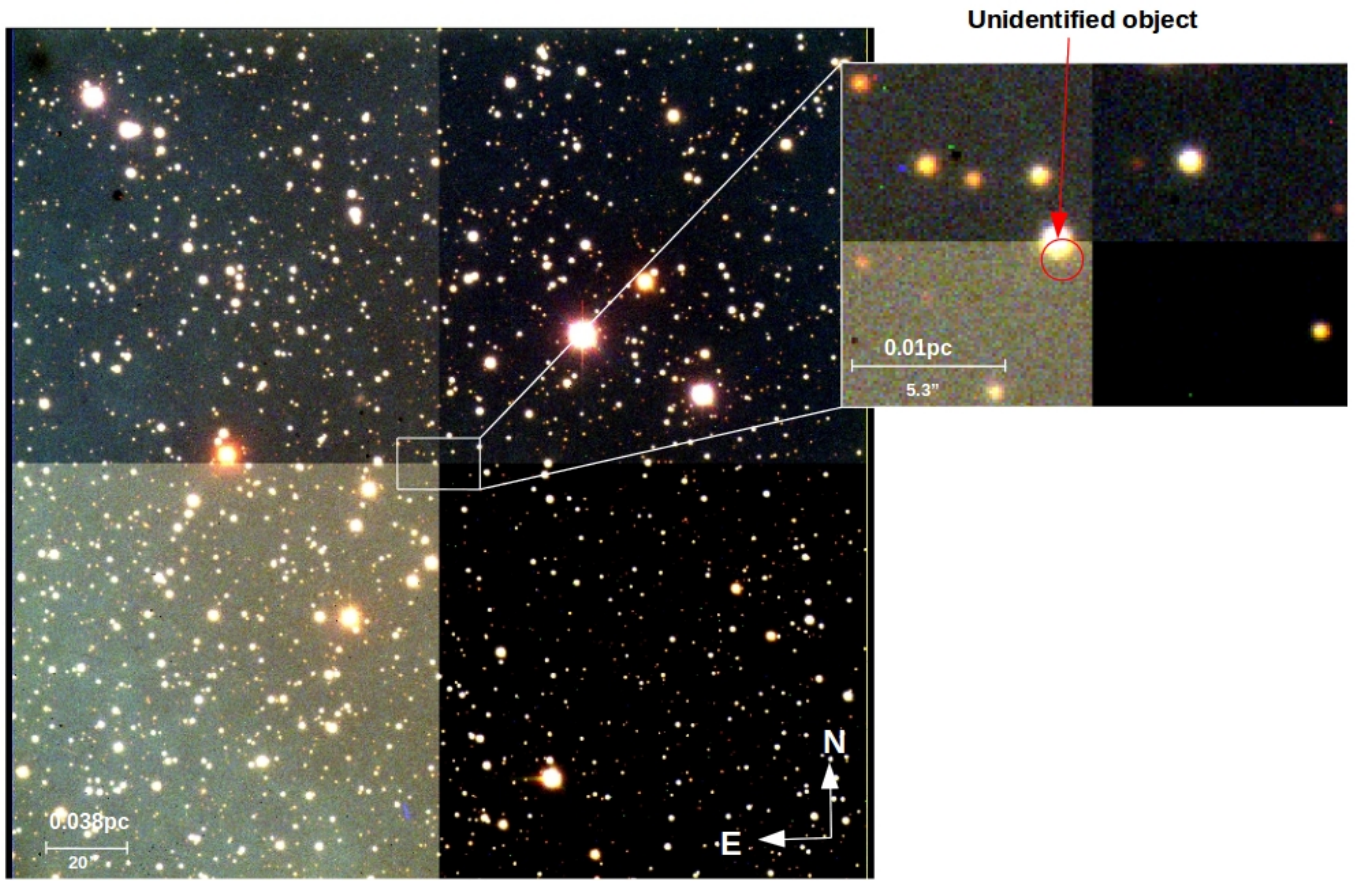

**Figure A10.** VLT VBR color composite image of PSR J1744-1134 showing the bright object near pulsar (marked in red).

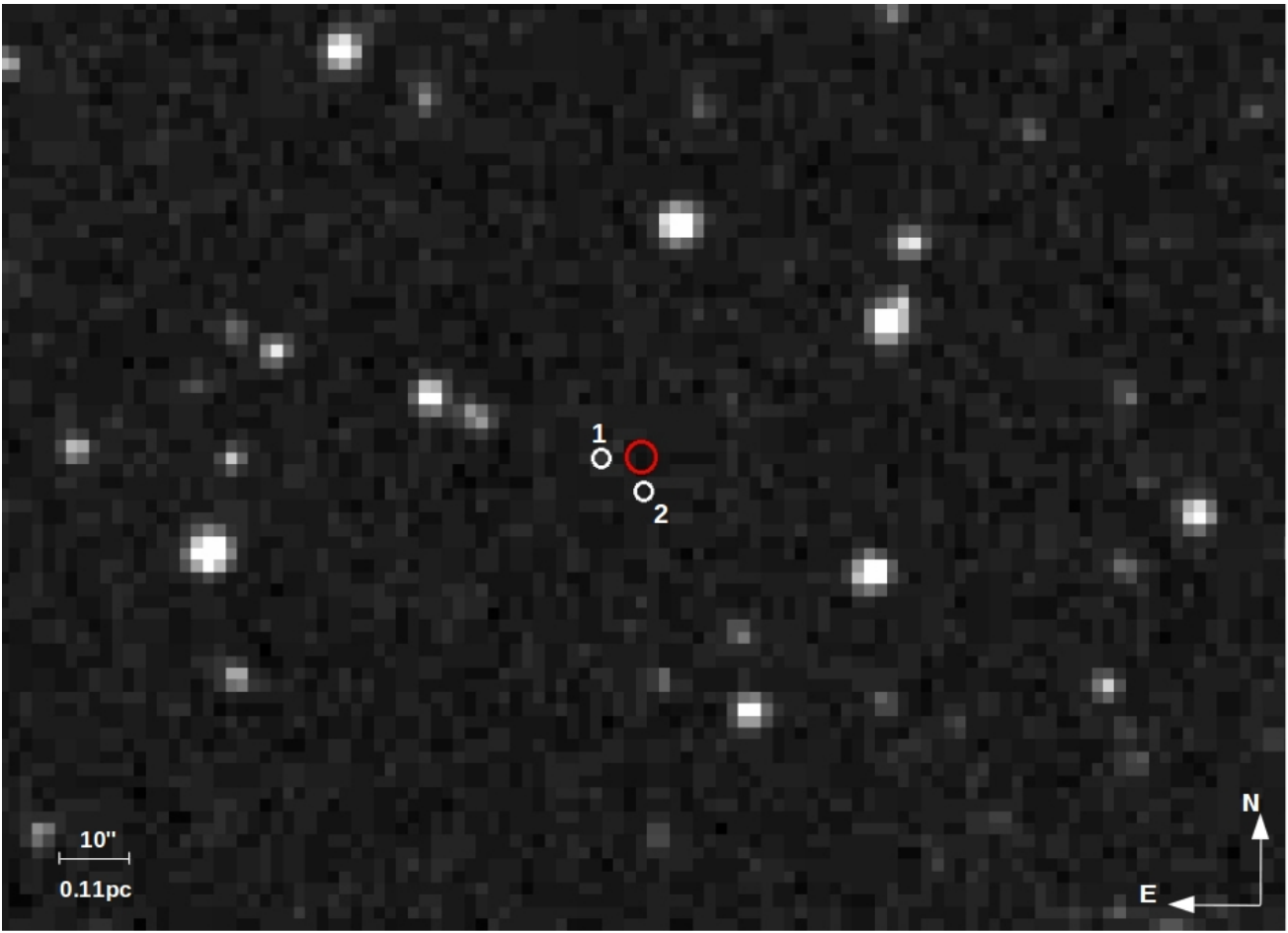

**Figure A11.** DSS image of PSR J1801-1417 with pulsar marked in red circle and the stars marked in white circles (1 and 2).

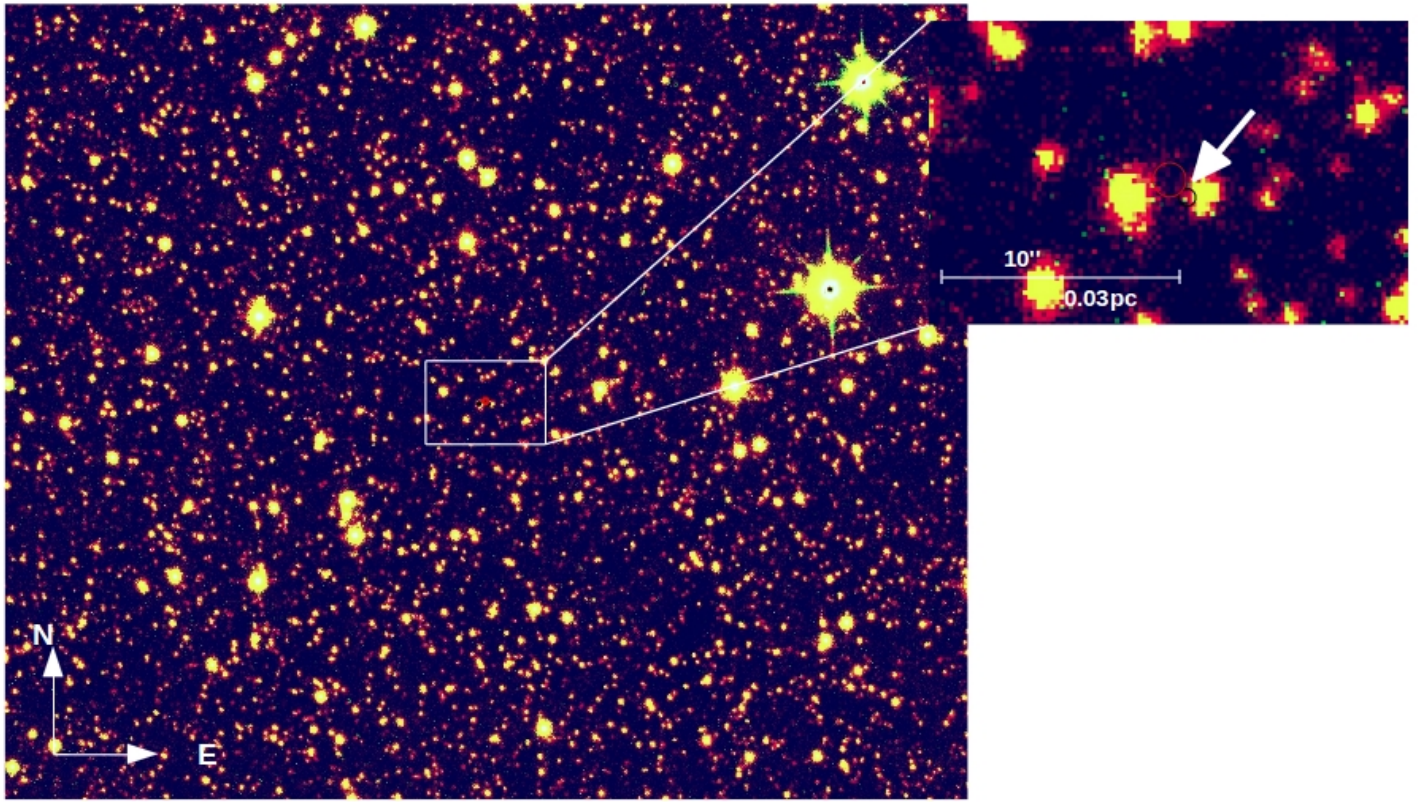

**Figure A12.** UKIDSS JHK color composite image of PSR J1843-1448 showing pulsar position marked in red circle and the star marked in black circle.

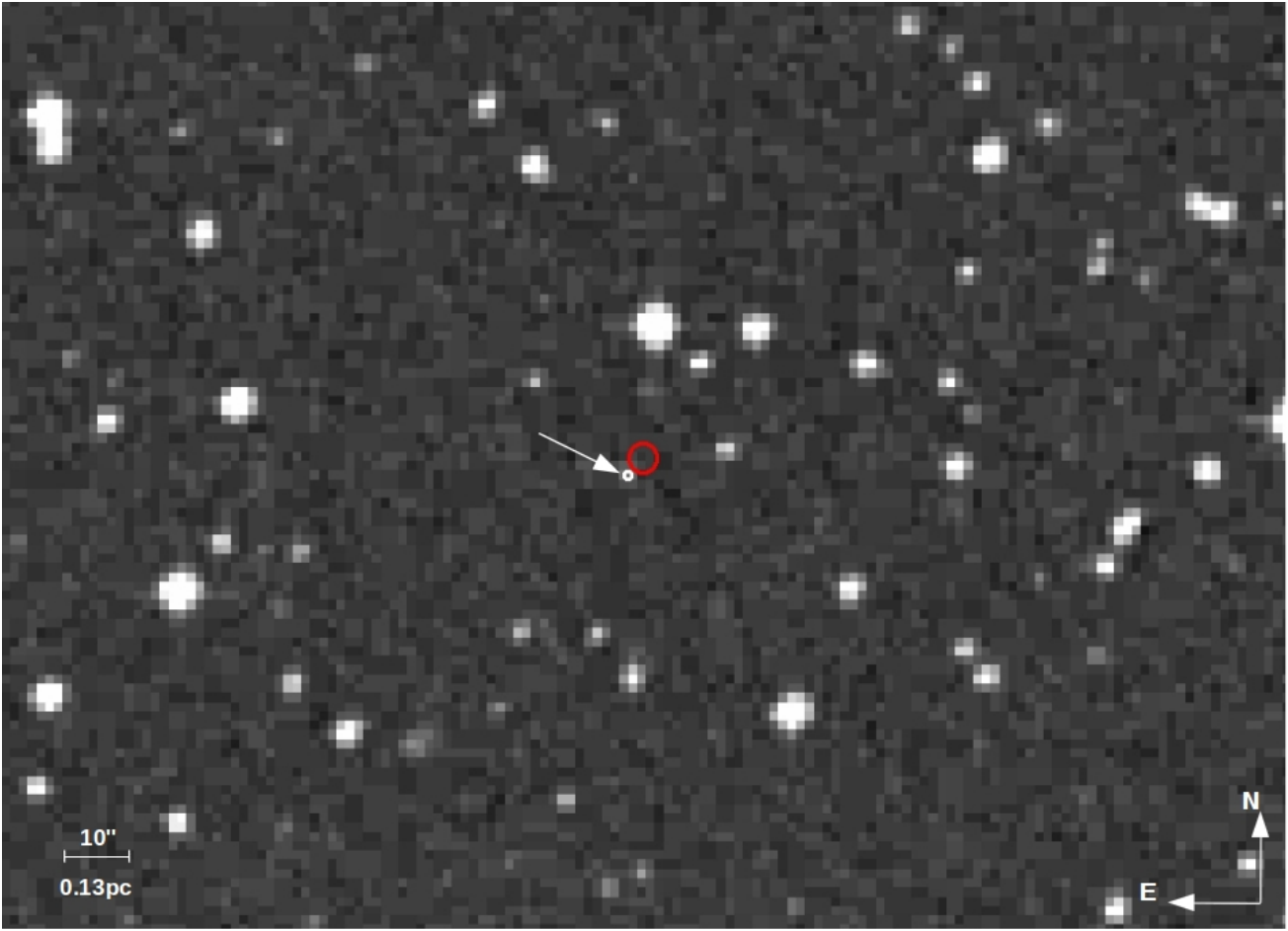

**Figure A13.** DSS image of PSR J1902-70 with pulsar marked in red circle and the star marked in white circle (arrow pointing).

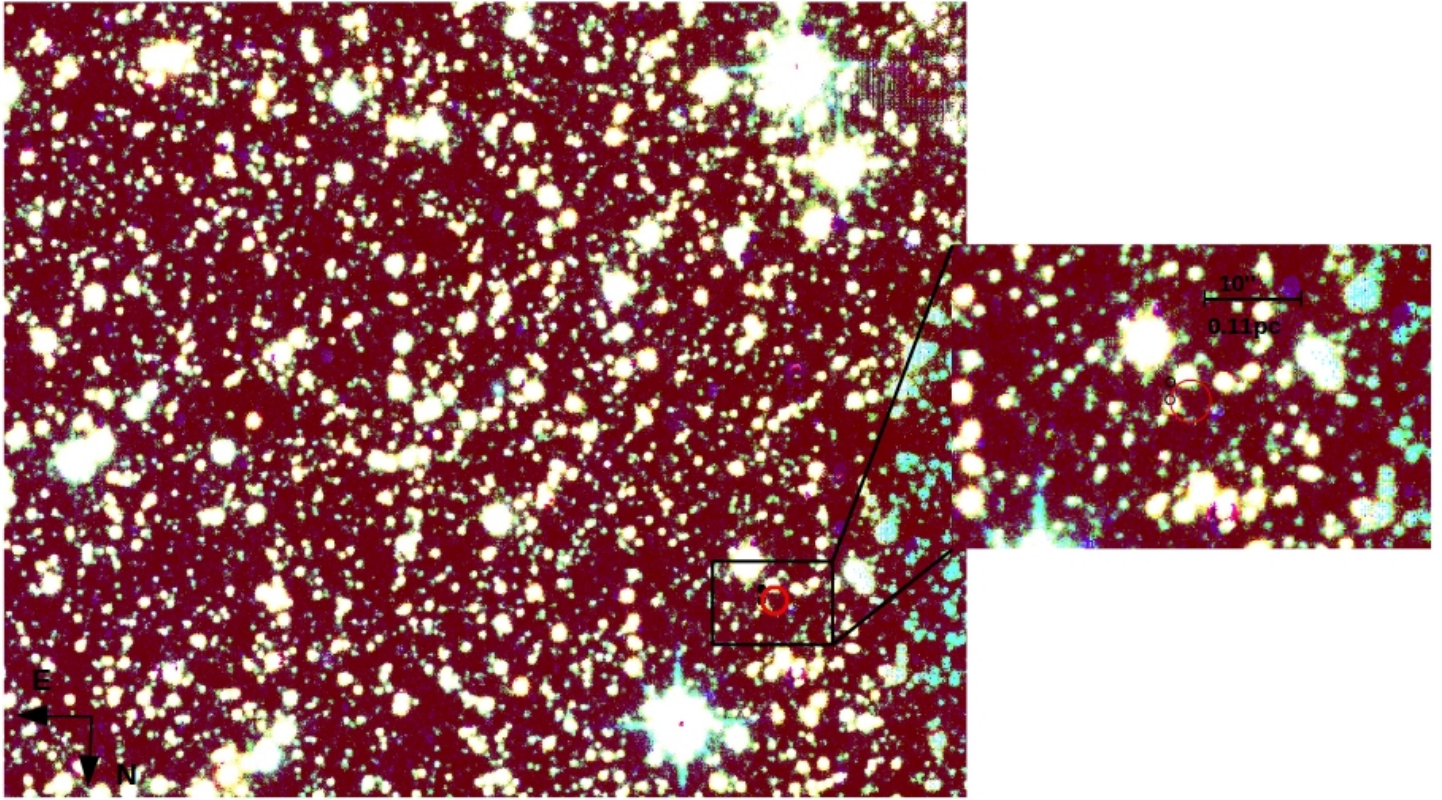

**Figure A14.** UKIDSS JHK color composite image of PSR J1905+0400 showing pulsar position marked in red circle and the stars marked in black circles.

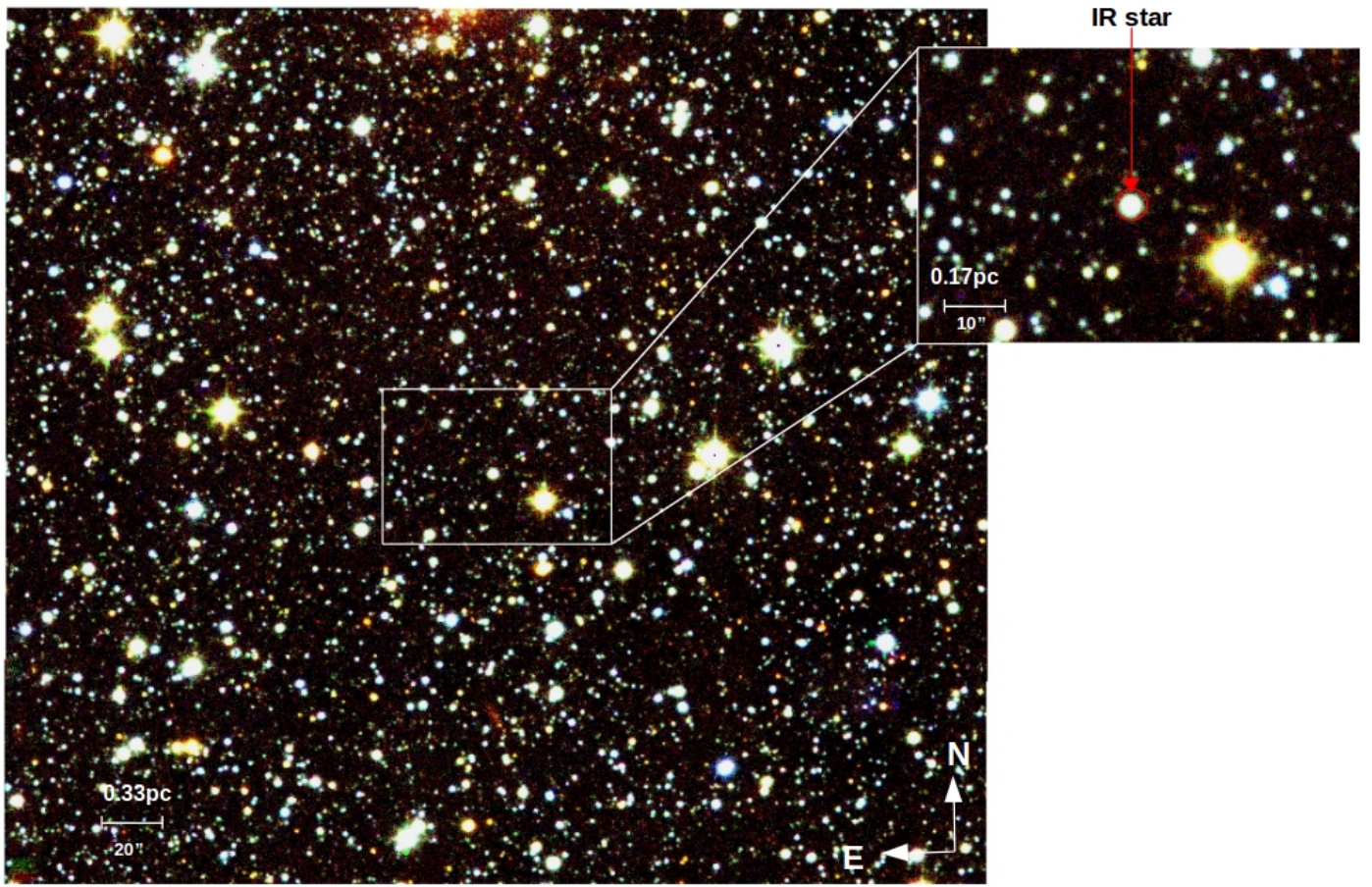

**Figure A15.** UKIDSS JHK color composite image showing PSR B1937+21 inside red circle and IR object at the same spot.